\documentclass[12pt,preprint]{aastex}

\shorttitle{Supermassive Black Holes in Galaxy Clusters}
\shortauthors{Fujita and Reiprich}

\begin{document}

\title{Can Supermassive Black Holes Sufficiently Heat 
Cool Cores of Galaxy Clusters?}

\author{Yutaka Fujita\altaffilmark{1,2},
and
Thomas H. Reiprich\altaffilmark{3}
}

\altaffiltext{1}{National Astronomical Observatory, Osawa 2-21-1,
Mitaka, Tokyo 181-8588, Japan; yfujita@th.nao.ac.jp}
\email{yfujita@th.nao.ac.jp}

\altaffiltext{2}{Department
of Astronomical Science, The Graduate University for Advanced Studies,
Osawa 2-21-1, Mitaka, Tokyo 181-8588, Japan}

\altaffiltext{3}{Department of Astronomy, University of
Virginia, P. O. Box 3818, Charlottesville, VA 22903-0818, USA;
thomas@reiprich.net}

\begin{abstract}
 Activities of a supermassive black hole or active galactic nucleus in
 the central galaxy of a cluster of galaxies have been promising
 candidates for heating sources of cool cluster cores. We estimate the
 masses of black holes using known correlations between the mass of a
 black hole and the velocity dispersion or the luminosity of the host
 galaxy. We find that the masses are $\sim 10^{8-9}\; M_\sun$ and the
 central X-ray luminosities of the host clusters (``the strength of the
 cooling flow'') are well below the Eddington luminosities. However, we
 do not find a correlation between the mass and the central X-ray
 luminosity of the host cluster. If the heating is stable, this seems to
 contradict a simple expectation if supermassive black holes are the
 main heating source of a cluster core. Moreover, if we assume a
 canonical energy conversion rate (10\%), black holes alone are unable
 to sufficiently heat the clusters with strong centrally peaked X-ray
 emission (``massive cooling flows'') over the lifetime of cluster
 cores.  These results may indicate that massive cooling flows are a
 transient phenomenon, which may be because the black holes are
 activated periodically. Alternatively, in the massive cooling flow
 clusters, the energy conversion rate may be larger than 10\%, that is,
 the black holes may be Kerr black holes.
\end{abstract}


\section{Introduction}

Clusters of galaxies are the largest gravitationally bound virialized
objects in the universe. They are filled with hot X-ray gas with
temperatures of $\sim 2$--$10$~keV. The gas is thought to be heated by
the energy released when clusters gravitationally collapsed. While the
gas cooling through the X-ray radiation is inefficient in most regions
of a cluster, it is very efficient in the central region because of the
high gas density. From X-ray observations, it can be estimated that the
thermal energy of the gas within cluster cores is lost radiatively
within $\sim 10^{8-9}$ yr. The gas cooling should decrease the
temperature and pressure of the gas at the cluster cores, and the weight
of the overlying gas should produce gas flows toward the cluster
centers. The flows are called ``cooling flows'' and the idea has
prevailed for about 30 years \citep{fab94}.

Recent X-ray observations showed that the gas in the central regions of
clusters is not cooling as much as previously believed. This is
indicated by the lack of metal line emission that is characteristic of
cooling X-ray gas
\citep[e.g.][]{ike97,mak01,pet01,tam01,kaa01,mat02}. This means that the
above simple cooling flow picture is not correct and that there must be
some heating sources balancing the energy loss through X-ray emission in
the central regions of clusters.

There are several heating models such as heat conduction
\citep[e.g.][]{tak81,tuc83,fri86,gae89,boh89a,spa92,sai99,nar01},
acoustic waves propagating from the outside of a cluster core
\citep*{fuj04}, or magnetohydrodynamic (MHD) effects \citep{mak01}. At
present, the most promising candidate for the heating source is the
supermassive black hole at the center of the giant elliptical
cluster-center galaxy
\citep*[e.g.][]{tuc83,rep87,boh88,bin95,sok01,cio01,boh02,
chu02,sok02,rey02,kai03}. In fact, the {\it Chandra} X-ray observatory
has shown that the strong activities of the supermassive black holes
disturb hot X-ray gas in the central regions of clusters
\citep*[e.g.][]{fab00,mcn00,bla01,mcn01,maz02,fuj02,joh02,kem02,tak03}. In
many clusters, bubbles of high energy particles have been found; it is
expected that jets produced by the supermassive black holes inflate
those bubbles. Moreover, numerical simulations have suggested that the
bubbles move outward in a cluster by buoyancy, thus mixing the
surrounding hot X-ray gas
\citep*[e.g.][]{chu01,qui01,sax01,bru02,bas03}. As a result of the
mixing, hot X-ray gas in the outer region of the cluster is brought
into, and subsequently heats the cluster center. Moreover, {\it Chandra}
found that acoustic waves formed through jet activities are propagating
in the hot X-ray gas of the Perseus cluster \citep{fab03}. Through the
viscosity of the gas, the waves may heat the surrounding gas.

The mass of a supermassive black hole generally needs to be large ($\sim
10^{9-10}\: M_\odot$) to balance the cooling of X-ray gas regardless of
the actual heating mechanism (e.g. bubbles, waves, and so on), if the
strong X-ray emission from the cluster is not a transient phenomenon
\citep*{fab02}. Moreover, relatively high heating efficiency may be
required. For example, assuming Bondi accretion, {\it XMM-Newton}
observations showed that the supermassive black hole at the center of
the Virgo cluster has to convert 5\% of the rest mass of the accreted
gas into thermal energy of the surrounding gas to prevent cooling at
least at present \citep[][see also \citealt{dim03}]{chu02}. This is
close to a canonical value that is often used in this kind of studies
($\sim 10$\%). In this paper, we derive the masses of supermassive black
holes and compare the maximum energy they can produce with the total
energy released as X-rays from the central regions of the clusters. In
this paper, we assume $h = 0.7$ ($H_0=100 h\:\rm km\: s^{-1}\:
Mpc^{-1}$), $\Omega_0=0.3$, and $\Lambda=0.7$ unless otherwise noted.

\section{Samples and Analysis}

\subsection{Black Hole Masses}

We study 53 clusters that have been observed with the {\it ROSAT}
satellite \citep{per98}, and belong to the X-ray flux limited sample of
55 clusters \citep*{edg92}. Unfortunately, clusters of galaxies are
generally distant, and the mass of the supermassive black hole at the
center of a cluster cD galaxy has been directly measured only for a few
clusters. However, it has been known that the mass of the black hole is
related to the properties of the host galaxy. Using {\it HST} and other
observations, the relations have been determined more exactly than early
studies \citep[e.g.][]{mag98}. In this paper, we use the relations
between the black hole and the host galaxy to obtain the black hole
mass. The mass of the supermassive black hole at the center of a cluster
cD galaxy can be obtained from empirical $M_{\rm BH}$--$\sigma$ and
$M_{\rm BH}$--$M_R$ relations, where $M_{\rm BH}$ is the mass of a
supermassive black hole, $\sigma$ is the velocity dispersion of stars in
the host galaxy, and $M_{\rm R}$ is the $R$-band absolute magnitude of
the host galaxy. For the $M_{\rm BH}$--$\sigma$ relation, we adopt
\begin{equation}
\label{eq:sigma}
 \log(M_{\rm BH}/M_\odot) = (8.12\pm 0.07) + (3.75\pm 0.3)
\log(\sigma/200\;\rm km\: s^{-1}) \;
\end{equation}
\citep{geb00}, and for the $M_{\rm BH}$--$M_R$ relation, we adopt
\begin{equation}
\label{eq:MR}
 \log(M_{\rm BH}/M_\odot) =  -(0.50\pm 0.02) M_R -(3.33\pm 0.48) \:
\end{equation}
\citep{mcl02}. The coefficients are slightly different from those in the
original papers because of differences in the assumed cosmological
parameters. There are more recent studies of the $M_{\rm BH}$--$\sigma$
relation that give steeper slopes; \citet{tre02} obtained 4.02 and
\citet{fer02} found 4.58. \citet{tre02} discussed the difference between
\citet{tre02} and \citet{fer02} and indicated that the large value
obtained by \citet{fer02} is due to inappropriate treatment of central
velocity dispersion. We confirmed that the difference between
\citet{geb00} and \citet{tre02} is so small that it does not affect the
results in this paper \citep[see Figure 26 in][]{pin03}. More recently,
\citet{pin03} studied the $M_{\rm BH}$--$\sigma$ relation only for
early-type galaxies and the slope they obtained is almost the same as
that obtained by \citet{geb00}, although the number of galaxies in the
sample is only 10. From these reasons, we use the result of
\citet{geb00}. The optical data for the galaxies at cluster centers,
such as the velocity dispersions and absolute magnitudes, are obtained
from the HYPERLEDA database \footnote{http://leda.univ-lyon1.fr/}. If
there are several measurements for a given cluster, we adopted the
latest one. Among the 53 clusters, the velocity dispersions of the
central galaxies are measured for 29 clusters. Since there are only 14
galaxies for which the $R$-band magnitudes are obtained, we use $B$-band
magnitudes for other galaxies; $B$-band magnitudes have been measured
for 40 galaxies including the above 14 galaxies. For the galaxies for
which only $B$-band magnitudes are obtained, the $B$-band magnitudes are
transformed to the $R$-band magnitudes using {\it $B$--$R$} color, where
the color is assumed to be the average of the 14 galaxies,
$B$--$R=1.67\pm 0.01$. The black hole masses are shown in
Table~\ref{tab:cluster}; the black hole mass derived from the $M_{\rm
BH}$--$\sigma$ relation is referred to as $M_{{\rm BH}, \sigma}$ and
that derived from the $M_{\rm BH}$--$M_R$ relation is referred to as
$M_{{\rm BH}, M_R}$. There are 43 clusters for which $M_{{\rm BH},
\sigma}$ or $M_{{\rm BH}, M_R}$ can be determined.  The errors in the
black hole masses are calculated using the worst case ranges given in
equations~(\ref{eq:sigma}) and~(\ref{eq:MR}).

\subsection{X-Ray Luminosities}

In addition to comparing current X-ray luminosities of clusters and the
masses of the central black holes, we compare the total energy released
in X-rays from the central region of clusters throughout the cluster
lifetime with the maximum energy that the black holes can supply. We use
the {\it ROSAT} data obtained by \citet{per98}. While data from {\it
Chandra} and {\it XMM-Newton} might be better suited for our purpose,
{\it ROSAT} data are the only ones available for a sufficiently large
number of clusters to conduct a statistically relevant uniform study.

The hot gas in a cluster core loses its thermal energy through X-ray
emission. Without an energy supply, the gas temperature would go to zero
within $\sim 10^{8-9}$~yr \citep{fab94}. The total energy that must be
supplied to prevent the cooling depends on the age of the cluster core
($t_{\rm age}$), which should be larger than that of the cluster
itself. In this paper, we assume that cluster cores formed at $z\sim 1$
because numerical simulations indicated that X-ray emission from a
cluster has not changed much since then \citep*{eke98}, and because a
significant part of the X-ray emission from a cluster comes from its
central region including the core. Note that the simulations done by
\citet{eke98} include neither cooling nor heating. The hot gas in the
central region of a cluster is generally in pressure equilibrium with
the gravity of dark matter because of the small sound crossing
time. Therefore, the simulations done by \citet{eke98} indicate that the
gravitational structure of cluster cores formed at $z\sim 1$. In actual
clusters, observations suggest that both cooling and heating are
effective but they appear to be well balanced \citep{pet01,tam01,kaa01}.
In fact, \citet*{all01} showed that temperature profiles among clusters
are similar and that the central temperatures drop only a factor of
$1/2$ even for ``cooling flow clusters'' \citep[see
also][]{ike01}. Although there is no consensus on the mechanism that
keeps the balance \citep[for thermal conduction, see][]{rus02,kim03},
the balanced cooling and heating may mean that the gas motion in the
central regions of clusters is slow and that the X-ray emission from
cluster cores reflects the gravitational structure there, which formed
at $z\sim 1$. Since there is no reason that $z=0$ is a special epoch, we
assume that the cooling and heating have been balanced since the
formation of the cluster core, and thus we assume that the X-ray
emission from the core has not changed much since $z\sim 1$. In fact,
\citet{val02} performed numerical simulations including both cooling and
heating and showed that the X-ray luminosity of a cluster has not
changed or even has decreased since $z\sim 1.2$ (high resolution
simulations in his Figure~11). The heating sources of the gas in these
simulations are supernovae, not supermassive black holes. However, since
the radiatively cooled gas is the fuel for both sources, the results
should be at least qualitatively the same. The reason for the decrease
in X-ray luminosity is that radiative cooling dominates heating before
enough stars are formed out of the cooled gas. After the cooling and
heating are balanced, the central gas density and X-ray luminosity of
the cluster decrease. Therefore, although we estimate the required
energy supply from a black hole, assuming that the X-ray luminosity of a
cluster core has been unchanged for $z\lesssim 1$, the supply may be
underestimated if we fix the X-ray luminosity at the observed value at
$z\sim 0$.

On the other hand, semi-analytic models of galaxy formation predict that
the supermassive black holes at galaxy centers gain most of their mass
at $z\lesssim 1$--$2$ \citep*{hae02,hat03,vol03}. Thus, we also assume
that the formation epoch of the supermassive black holes is $z\sim 1$.

Since we assume that cluster cores formed at $z\sim 1$, the hot X-ray
gas within radii where the cooling time of the gas is $t_{\rm
age}=7.7$~Gyr (the look-back time for $z=1$) must be
heated. \citet{per98} listed the X-ray luminosities of 53 clusters
within radii where the cooling time of hot X-ray gas is 13~Gyr. In a
classical cooling flow model, the central X-ray luminosity is correlated
with the strength of the cooling flow \citep{fab94}. We modify the
central X-ray luminosities listed by \citet{per98} into those for the
cooling time of 7.7~Gyr to estimate the energy supplied by supermassive
black holes.

In order to modify the central X-ray luminosity, we need to find the
density profile of the X-ray gas in a cluster. We assume that a cluster
is isothermal. For ``cooling flow clusters,'' {\it Chandra} observations
showed that the temperature at the center is $\sim 1/2$ of the average
\citep{all01}. However, this temperature drop does not affect photon
counts in the {\it ROSAT} band \citep[at most $\sim 10$\%,
see][]{moh99}. A parametric description of the cluster gas density
profile has often been used, and it is called the $\beta$-model
\citep[e.g.][]{cav76}. Assuming spherical symmetry, the model
\begin{equation}
\label{eq:surf}
 S_X(R) = S_0 \left(1+\frac{R^2}{r_c^2}\right)^{-3\beta+1/2}
\end{equation}
is fitted to the measured surface brightness profile, where $R$ is the
projected distance from the cluster center.  This yields values for the
core radius, $r_c$, the parameter, $\beta$, and the normalization,
$S_0$. From the fit values, the radial gas density distribution can be
constructed ($\beta$ model);
\begin{equation}
\label{eq:rhogas_ob}
 \rho_{\rm gas}(r)=\rho_{\rm gas, 0} 
\left(1+\frac{r^2}{r_c^2}\right)^{-3\beta/2} \:.
\end{equation}

If excess emission (so-called cooling flow) is seen in the innermost
region of a cluster, the surface profile can be fitted with two
$\beta$-model functions
\begin{equation}
\label{eq:surf1}
 S_X(R) = S_{01} \left(1+\frac{R^2}{r_{c1}^2}\right)^{-3\beta_1+1/2}+
S_{02} \left(1+\frac{R^2}{r_{c2}^2}\right)^{-3\beta_2+1/2} \:,
\end{equation}
where $r_{c1}<r_{c2}$ and $S_{01}>S_{02}$.  In this case, the density
profile is given by
\begin{equation}
\label{eq:rhogas_ob1}
 \rho_{\rm gas}(r)^2=\rho_{\rm gas, 0, 1}^2 
\left(1+\frac{r^2}{r_{c1}^2}\right)^{-3\beta_1} 
+ \rho_{\rm gas, 0, 2}^2 
\left(1+\frac{r^2}{r_{c2}^2}\right)^{-3\beta_2} \:,
\end{equation}
where $\rho_{\rm gas, 0, 1}>\rho_{\rm gas, 0, 2}$.

The results of the fits are shown in Table~\ref{tab:beta}; the results
of single-$\beta$-model fits are represented by subscript~2. Among 106
clusters investigated by \citet{rei02}, there are 24 clusters that
overlap with the 53 clusters studied by \citet{per98} and that the
qualities of the data are high enough. For these clusters, we fit the
surface brightness with two $\beta$ models without the assumption of
$\beta_1=\beta_2$ and obtain the density profile
(equation~[\ref{eq:rhogas_ob1}]). The selection criteria of the clusters
with high data quality are (1) {\it ROSAT} PSPC pointed observations had
been made, and (2) $S_{02}>0$ and $S_{01}>S_{02}$. These clusters are
indicated by `RB2' in Table~\ref{tab:beta}.

Among the remaining, 22 clusters were studied by \citet*{moh99}. For the
clusters with central excess, they fit the surface brightness and
obtain density profiles assuming that $\beta_1 = \beta_2$. Since they
presented only the central gas density, $\rho_{\rm gas, obs, 0}$, for
the clusters with a central excess, we estimate $\rho_{\rm gas, 0, 1}$
and $\rho_{\rm gas, 0, 2}$ by solving equations
\begin{equation}
 \rho_{\rm gas, 0, 1}^2+\rho_{\rm gas,
0, 2}^2 = \rho_{\rm gas, obs, 0}^2 \:,
\end{equation}
\begin{equation}
 \label{eq:rho1}
 \rho_{\rm gas, 0, 2} = 
\left(\frac{S_{02} r_{c1}}{S_{01} r_{c2}}\right)^{1/2}
\rho_{\rm gas, 0, 1} \:.
\end{equation}
For the Virgo cluster, we use the result of a double-$\beta$-model fit
done by \citet{mat02}. For the rest clusters, we use the results of
single-$\beta$-model fits obtained by \citet{rei02}. These clusters are
indicated by `RB1' in Table~\ref{tab:beta}.

Unfortunately, $\beta$-model fits have not yet been studied
systematically with {\it Chandra} and {\it XMM-Newton}. However, for the
Coma cluster, for example, \citet{neu03} determined that
$r_c=253$--340~kpc and $\beta=0.72$--0.85, which is not much different
from the results obtained with {\it ROSAT} (Table~\ref{tab:beta}).

\citet{per98} defined a cooling radius, $r_{\rm cool, P}$, as the radius
at which the cooling time of X-ray gas is $t_{\rm cool}=13$~Gyr for
their cosmological parameters ($h=0.5$, $\Omega_0=1$, and $\Lambda=0$)
and presented derived $r_{\rm cool, P}$ in their Table~5. Below, we use
the PSPC data in \citet{per98} if a cluster was observed by both PSPC
and HRI. Since the cooling time of X-ray gas is proportional to
$\rho_{\rm gas}^{-1}$ for a given temperature, and since $\rho_{\rm
gas}\propto h^{1/2}$, $t_{\rm cool}=13$~Gyr in \citet{per98} corresponds
to $t_{\rm cool}=13/\hat{h}^{1/2}\approx 11$~Gyr for our cosmological
parameters, where $\hat{h}=1.4$ is the ratio of the Hubble constant we
assumed to that of \citet{per98}. (In actual calculations, we consider
the effect of $\Omega_0\neq 1$. Thus $\hat{h}$ is not exactly 1.4.)
Thus, assuming that a cluster is isothermal at $r\sim r_{\rm cool, P}$,
the radius at which $t_{\rm cool}=7.7$~Gyr ($=t_{\rm age}$) must satisfy
the relation:
\begin{equation}
 \frac{\rho_{\rm gas}(r[t_{\rm cool}=7.7\rm\; Gyr])}
{\rho_{\rm gas}(r[t_{\rm cool}=13\hat{h}^{-1/2} \rm\; Gyr])}
=\frac{13\: \hat{h}^{-1/2}}{7.7}\;.
\end{equation}
Here $r(t_{\rm cool}=13\:\hat{h}^{-1/2}\rm\; Gyr)$ corresponds to
$r_{\rm cool, P}/\hat{h}$. We define $r_{\rm cool, FR}\equiv r(t_{\rm
cool}=7.7\rm\; Gyr)$ to be discriminated from $r_{\rm cool, P}$.

\citet{per98} also derived X-ray luminosities of clusters within their
cooling radii, $L_{\rm P}(<r_{\rm cool, P})$, for their cosmological
parameters (their Table~5). The X-ray luminosity of a cluster within
$r_{\rm cool, FR}$ for our cosmological parameters is estimated by
\begin{equation}
 L_{\rm FR}(<r_{\rm cool, FR}) 
= \frac{\int_0^{r_{\rm cool, FR}}\rho_{\rm gas}(r)^2 r^2 dr}
  {\int_0^{r_{\rm cool, P}/\hat{h}}\rho_{\rm gas}(r)^2 r^2 dr}
L_{\rm P}(<r_{\rm cool, P})\hat{h}^{-2}\;.
\end{equation}
In Table~\ref{tab:cluster}, we present $L_{\rm FR}(<r_{\rm cool,
FR})$. The modified luminosity, $L_{\rm FR}(<r_{\rm cool, FR})$, is
always smaller than $L_{\rm P}(<r_{\rm cool, P})$. 

\section{Discussion}

\begin{figure}\epsscale{0.45}
\plotone{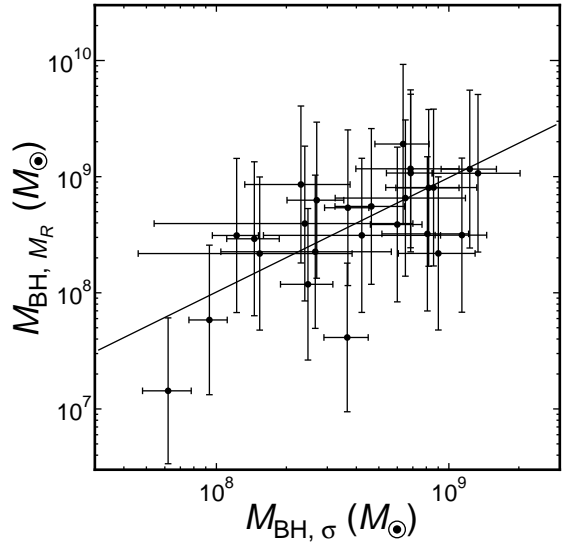} \caption{Black hole mass derived from the $M_{\rm
BH}$--$\sigma$ relation ($M_{{\rm BH}, \sigma}$) versus that derived
from the $M_{\rm BH}$--$M_R$ relation ($M_{{\rm BH}, M_R}$) for 26
clusters for which both velocity dispersion and absolute magnitude are
measured. The line indicates $M_{{\rm BH}, M_R}=M_{{\rm BH},
\sigma}$.}\label{fig:MsigMR}
\end{figure}

Contrary to normal elliptical galaxies, the central galaxies of clusters
often have diffuse optical envelopes; thus, the absolute magnitude
depends on the radius within which observers measure the galactic
luminosity. Unfortunately, the radii are not defined for the galaxies
listed at HYPERLEDA, which increases the overall uncertainty. Thus, as
the mass of a black hole ($M_{\rm BH}$), we adopt that derived from the
$M_{\rm BH}$--$\sigma$ relation ($M_{{\rm BH}, \sigma}$) rather than
that derived from the $M_{\rm BH}$--$M_R$ relation ($M_{{\rm BH}, M_R}$)
for 26 clusters for which both velocity dispersion and $B$ or $R$-band
absolute magnitude of the central galaxy are obtained. We note, however,
that there is no systematic difference between $M_{{\rm BH}, \sigma}$
and $M_{{\rm BH}, M_R}$ (Figure~\ref{fig:MsigMR}). In general the error
bars of $M_{{\rm BH}, M_R}$ are much larger than those of $M_{{\rm BH},
\sigma}$, which blurs the correlation (Figure~\ref{fig:MsigMR}). If only
$M_{{\rm BH}, \sigma}$ (or $M_{{\rm BH}, M_R}$) is obtained, we call it
$M_{\rm BH}$.

\begin{figure}\epsscale{0.45}
\plotone{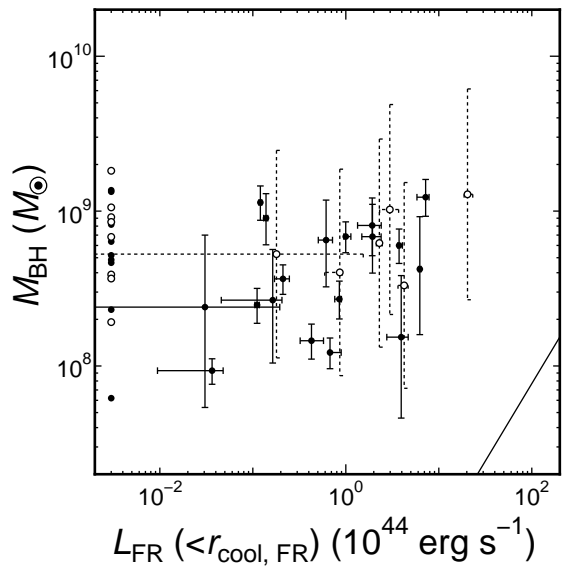} \caption{The central X-ray luminosity of a cluster,
 $L_{\rm FR}(<r_{\rm cool, FR})$, versus the mass of the black hole at
 the center of the central galaxy, $M_{\rm BH}$. The black hole masses
 are determined by the $M_{\rm BH}$--$\sigma$ relation (filled circles
 with solid errors) or the $M_{\rm BH}$--$M_R$ relation (open circles
 with dotted errors). Data points of the clusters with $L_{\rm
 FR}(<r_{\rm cool, FR})=0$ are shown on the left side of the figure. The
 line indicates the Eddington luminosity, $L_{\rm Edd}$.}\label{fig:LM}
\end{figure}

Figure~\ref{fig:LM} shows the central X-ray luminosity of a cluster
versus the mass of the supermassive black hole. As can be seen, there is
no correlation between these parameters.  In fact, a Spearman rank
coefficient is almost zero ($r_s=4\times 10^{-3}$), which means no
correlation. This seems to contradict the idea that a supermassive black
hole is the main heating source of the cluster core; if the supermassive
black hole were the sole heating source, one would expect a larger
central X-ray luminosity for a larger black hole mass. In
Figure~\ref{fig:LM}, we present the Eddington luminosity:
\begin{equation}
 L_{\rm Edd} = 1.3\times 10^{46} (M_{\rm BH}/10^8 M_\sun) 
\;\rm erg\; s^{-1}
\end{equation}
\citep{sha83}. Since $L_{\rm FR}<L_{\rm Edd}$, the energy injection
rates from black holes at present do not need to be super-Eddington or
unrealistically high.

\begin{figure}\epsscale{0.45}
\plotone{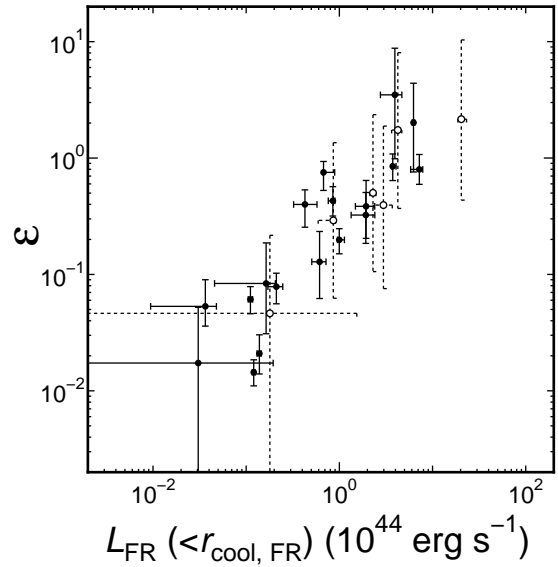} \caption{The central X-ray luminosity of a cluster,
 $L_{\rm FR}(<r_{\rm cool, FR})$, versus the energy conversion rate,
 $\varepsilon$. Black hole masses are determined by the $M_{\rm
 BH}$--$\sigma$ relation (filled circles with solid errors) or the
 $M_{\rm BH}$--$M_R$ relation (open circles with dotted
 errors).}\label{fig:LR}
\end{figure}

If we assume a canonical radiative efficiency of $\eta=0.1$, the total
energy that a supermassive black hole can release is $\eta M_{\rm
BH}c^2$, where $c$ is the velocity of light. On the other hand, $L_{\rm
FR}(<r_{\rm cool, FR}) t_{\rm age}$ is the energy that must be supplied
by the supermassive black hole if $L_{\rm FR}(<r_{\rm cool, FR})$ is
constant. Thus, the ratio $\varepsilon\equiv L_{\rm FR}(<r_{\rm cool,
FR}) t_{\rm age}/(\eta M_{\rm BH}c^2)$ is the necessary energy
conversion rate of the matter accreted onto a supermassive black hole
into heating of the X-ray gas in a cluster core. Figure~\ref{fig:LR}
shows that $\varepsilon\gtrsim 1$ for $L_{\rm FR}(<r_{\rm cool,
FR})\gtrsim 3\times 10^{44}\rm\; erg\; s^{-1}$, which means that a
supermassive black hole may not prevent gas cooling by X-ray emission
throughout the age of the cluster core, but note the large
uncertainty. These clusters are so-called ``massive cooling flow''
clusters with the mass deposition rates (expected in the absence of
heating) of $\dot{M}\gtrsim 200\; M_\sun\;\rm yr^{-1}$ (for $h=0.5$,
which has often been used in this field).

There are two possibilities to overcome this difficulty. One is that the
radiative efficiency $\eta$ is larger than the canonical value
(0.1). This happens if the black holes are rapidly rotating Kerr black
holes. In this case, the maximum value of $\eta$ is 0.42, based on the
binding energy of particles on the innermost stable circular orbit
\citep{sha83}. Moreover, Kerr holes also alter the energy that can be
stored and extracted from the black holes (e.g. via the Blandford-Znajek
mechanism). The other possibility is that a massive cooling flow or
strong X-ray emission from a cluster core is a transient event. This may
also be consistent with the fact that there is no correlation between
the mass of a black hole and the X-ray luminosity of a cluster
(Figure~\ref{fig:LM}), which may indicate that the cooling and heating
at the cluster center are not exactly balanced
\citep[see][]{sok01}. Here, we assume that the supermassive black holes
we selected are an unbiased sample of the whole supermassive black hole
population, and that the average of the energy conversion rates,
$\varepsilon$, can be regarded as the time-average of the conversion
rate for {\it a black hole}. However, since our cluster sample is
flux-limited, the bright clusters tend to be preferentially selected. In
order to correct for this effect, we take the average of $\varepsilon$
by weighting with $1/V_{\rm max}$, where $V_{\rm max}$ is the maximum
volume within which clusters with a given luminosity can be
observed. The volume is represented by $V_{\rm max}\propto L_X^{3/2}$,
where $L_X$ is the total X-ray luminosity of a cluster obtained from
\citet{dav93}. For 43 clusters for which we could estimate the mass of
supermassive black holes, the average conversion rate is $\langle
\varepsilon \rangle=6.4^{+4.7}_{-1.6}\times 10^{-2}$. Moreover, we
describe below how we correct the bias attributed to the 10 clusters for
which we could not estimate the black hole masses. The X-ray luminosity
and the back hole mass of a cluster are not correlated
(Figure~\ref{fig:LM}), and the average black hole mass for the 43
clusters is $6.3\times 10^8\; M_\sun$. Assuming that the black hole mass
for the remaining 10 clusters is the same as the average mass, we can
estimate $\varepsilon$ for these clusters. If we include these 10
clusters, the average of the conversion rate is $\langle \varepsilon
\rangle=6.6^{+4.7}_{-1.6}\times 10^{-2}$. Since $\langle \varepsilon
\rangle$ is much smaller than one, the supermassive black holes can heat
the surrounding hot X-ray gas of the clusters unless the past X-ray
luminosities were $\gtrsim 20$ times larger than the current
values. Some of the results of numerical simulations by \citet{val02}
show that the X-ray luminosity of a cluster is increasing toward higher
redshift (high resolution simulations in his Figure~11). However, the
increase is at most a factor of 10.

There is one more possibility; the $M_{\rm BH}$--$\sigma$ and $M_{\rm
BH}$--$M_R$ relations we used may not apply to the supermassive black
holes in the cD galaxies in clusters. As far as we know, there are two
clusters for which the masses of the supermassive black holes have been
directly measured. One is the Virgo cluster and the other is
Cygnus~A. The mass of the supermassive back hole at the center of the
Virgo cluster is $(3.2\pm 0.9)\times 10^9\; M_\sun$ \citep{mac97} and
that of Cygnus~A is $(2.5\pm 0.7)\times 10^9\; M_\sun$ \citep{tad03}. On
the other hand, the masses obtained from the $M_{\rm BH}$--$\sigma$
relation are $3.1_{-2.5}^{+11.3} \times 10^8 \; M_\sun$ and
$3.3_{-2.6}^{+12.0}\times 10^8 \; M_\sun$, respectively
(Table~\ref{tab:cluster}). Thus, the values obtained from the $M_{\rm
BH}$--$\sigma$ relation are significantly smaller than those directly
measured.  However, since there are only two samples, it is premature to
conclude that the discrepancy generally exists for cD galaxies.

\section{Conclusions}

We estimated the masses of supermassive black holes, $M_{\rm BH}$, at
the centers of 43 clusters of galaxies included in an X-ray flux limited
sample. We showed that $M_{\rm BH}\sim 10^{8-9}\; M_\sun$ using an
empirical relation between $M_{\rm BH}$ and the velocity dispersion (or
the luminosity) of the host galaxy. The central X-ray luminosities of
the host clusters are well below the Eddington luminosities. We found
that there is no correlation between $M_{\rm BH}$ and the central X-ray
luminosity of the cluster core. This seems to contradict a simple
expectation if a supermassive black hole is the main heating source of
the cluster core. We also showed that strong X-ray emission observed in
some cluster cores may not be sustained by the heating by supermassive
black holes for the ages of the cluster cores. These results may
indicate that the strong X-ray emission is a transient phenomenon, which
may be because the black hole activities are periodic. Moreover, the
clusters with the strong X-ray emission may be very effectively heated
by Kerr black holes.

\acknowledgments

We thank the anonymous referee for useful suggestions.  We also thank
T.~E. Clarke, I. Tanaka, Y. Sato, M. Machida, K. Shimasaku, T. Goto, M.,
Enoki, M. Nagashima, S. Iwamoto, N. Yoshida, K. Asano, and M. Kino for
useful comments. Y.~F.\ was supported in part by a Grant-in-Aid from the
Ministry of Education, Culture, Sports, Science, and Technology of Japan
(14740175).  T.~H.~R. acknowledges support by the Celerity Foundation
through a postdoctoral fellowship.

\clearpage

\begin{deluxetable}{cccccc}
\tabletypesize{\scriptsize}
\tablecaption{Cluster and Black Hole Parameters. \label{tab:cluster}}
\tablewidth{0pt}
\tablehead{
\colhead{Cluster} & \colhead{$\sigma$}   & \colhead{$M_R$}   &
\colhead{$M_{{\rm BH}, \sigma}$} &
\colhead{$M_{{\rm BH}, M_R}$}  & 
\colhead{$L_{\rm FR}(<r_{\rm cool, FR})$} \\
\colhead{} & \colhead{($\rm km\; s^{-1}$)}   & \colhead{}   &
\colhead{($10^8\; M_\sun$)} &
\colhead{($10^8\; M_\sun$)}  & 
\colhead{($10^{44}\rm\; erg\; s^{-1}$)} 
}
\startdata
A85           &
$322^{+  32}_{ -32}$&
$-22.9 $\tablenotemark{a}&
$  8.1^{+  4.1}_{ -2.9}$&
$  3.2^{+ 11.6}_{ -2.5}$&
$  1.9^{+  0.5}_{ -0.6 }$
 \\
A119          &
$278^{+  21}_{ -21}$&
$-23.4$&
$  4.6^{+  1.8}_{ -1.4}$&
$  5.5^{+ 20.4}_{ -4.4}$&
 0
 \\
A262          &
$236^{+  12}_{ -12}$&
$-22.1 $\tablenotemark{a}&
$  2.5^{+  0.7}_{ -0.6}$&
$  1.2^{+  4.1}_{ -0.9}$&
$ 0.11^{+ 0.01}_{-0.01 }$
 \\
AWM7          &
$333^{+  27}_{ -27}$&
$-22.6$&
$  9.0^{+  4.0}_{ -2.9}$&
$  2.2^{+  7.8}_{ -1.7}$&
$ 0.14^{+ 0.01}_{-0.01 }$
 \\
A399          &
$230^{+  30}_{ -30}$&
$-23.8 $\tablenotemark{a}&
$  2.3^{+  1.4}_{ -1.0}$&
$  8.6^{+ 32.0}_{ -6.8}$&
 0
 \\
A401          &
$367^{+  35}_{ -35}$&
$-24.0 $\tablenotemark{a}&
$ 13.3^{+  6.9}_{ -4.9}$&
$ 10.7^{+ 40.4}_{ -8.5}$&
 0
 \\
A3112         &
\nodata &
$-23.9 $\tablenotemark{a}&
\nodata &
$ 10.2^{+ 38.6}_{ -8.1}$&
$  3.0^{+  0.7}_{ -0.7 }$
 \\
A426          &
$272^{+  61}_{ -61}$&
$-22.9$&
$  4.2^{+  5.0}_{ -2.6}$&
$  3.1^{+ 11.3}_{ -2.4}$&
$  6.2^{+  0.2}_{ -0.2 }$
 \\
2A~0335$+$096 &
\nodata &
$-23.5 $\tablenotemark{a}&
\nodata &
$  6.2^{+ 23.0}_{ -4.9}$&
$  2.3^{+  0.2}_{ -0.2 }$
 \\
A3158         &
\nodata &
$-23.1 $\tablenotemark{a}&
\nodata &
$  3.9^{+ 14.2}_{ -3.1}$&
 0
 \\
A478          &
\nodata &
\nodata &
\nodata &
\nodata &
$  7.1^{+  0.9}_{ -1.3 }$
 \\
A3266         &
$327^{+  34}_{ -34}$&
$-23.7 $\tablenotemark{a}&
$  8.6^{+  4.6}_{ -3.2}$&
$  8.1^{+ 30.1}_{ -6.4}$&
 0
 \\
A496          &
$241^{+  14}_{ -14}$&
$-23.5 $\tablenotemark{a}&
$  2.7^{+  0.8}_{ -0.7}$&
$  6.3^{+ 23.2}_{ -4.9}$&
$ 0.85^{+ 0.04}_{-0.09 }$
 \\
A3391         &
\nodata &
$-24.0 $\tablenotemark{a}&
\nodata &
$ 10.6^{+ 39.8}_{ -8.4}$&
 0
 \\
A576          &
$282^{+  11}_{ -11}$&
\nodata &
$  4.8^{+  1.3}_{ -1.1}$&
\nodata &
 0
 \\
PKS~0745$-$191&
\nodata &
$-24.1 $\tablenotemark{a}&
\nodata &
$ 12.8^{+ 48.7}_{-10.2}$&
$ 20.4^{+  2.8}_{ -1.6 }$
 \\
A644          &
\nodata &
$-23.4 $\tablenotemark{a}&
\nodata &
$  5.3^{+ 19.4}_{ -4.1}$&
$ 0.18^{+ 1.36}_{-0.18 }$
 \\
A754          &
$323^{+  19}_{ -19}$&
$-23.7 $\tablenotemark{a}&
$  8.2^{+  2.9}_{ -2.3}$&
$  8.0^{+ 29.9}_{ -6.3}$&
 0
 \\
HYD-A         &
$308^{+  38}_{ -38}$&
$-24.1 $\tablenotemark{a}&
$  6.8^{+  4.2}_{ -2.9}$&
$ 11.7^{+ 44.2}_{ -9.2}$&
$  1.9^{+  0.5}_{ -0.4 }$
 \\
A1060         &
$182^{+   4}_{  -4}$&
$-21.5$&
$  0.9^{+  0.2}_{ -0.2}$&
$  0.6^{+  2.0}_{ -0.5}$&
$ 0.04^{+ 0.01}_{-0.03 }$
 \\
A1367         &
$163^{+   7}_{  -7}$&
$-20.2 $\tablenotemark{a}&
$  0.6^{+  0.2}_{ -0.1}$&
$  0.1^{+  0.5}_{ -0.1}$&
 0
 \\
Virgo         &
$355^{+   8}_{  -8}$&
$-22.9$&
$ 11.4^{+  3.2}_{ -2.6}$&
$  3.1^{+ 11.3}_{ -2.5}$&
$ 0.12^{+ 0.00}_{- 0.00 }$
 \\
Cent          &
$262^{+   8}_{  -8}$&
$-21.2$&
$  3.7^{+  0.8}_{ -0.8}$&
$  0.4^{+  1.4}_{ -0.3}$&
$ 0.21^{+ 0.04}_{-0.04 }$
 \\
Coma          &
$262^{+   8}_{  -8}$&
$-23.4$&
$  3.7^{+  0.8}_{ -0.8}$&
$  5.4^{+ 19.9}_{ -4.2}$&
 0
 \\
A1644         &
\nodata &
$-24.4 $\tablenotemark{a}&
\nodata &
$ 18.2^{+ 70.0}_{-14.5}$&
 0
 \\
A3532         &
\nodata &
\nodata &
\nodata &
\nodata &
 0
 \\
A1650         &
\nodata &
\nodata &
\nodata &
\nodata &
$ 0.00^{+ 1.90}_{- 0.00 }$
 \\
A1651         &
\nodata &
\nodata &
\nodata &
\nodata &
$ 0.15^{+ 0.57}_{-0.15 }$
 \\
A1689         &
\nodata &
\nodata &
\nodata &
\nodata &
$ 11.3^{+  5.2}_{ -0.9 }$
 \\
A1736         &
\nodata &
$-23.0 $\tablenotemark{a}&
\nodata &
$  3.7^{+ 13.3}_{ -2.9}$&
 0
 \\
A3558         &
\nodata &
$-23.8$&
\nodata &
$  9.1^{+ 34.3}_{ -7.2}$&
 0
 \\
A3562         &
$233^{+  76}_{ -76}$&
$-23.1 $\tablenotemark{a}&
$  2.4^{+  4.6}_{ -1.9}$&
$  4.0^{+ 14.4}_{ -3.1}$&
$ 0.03^{+ 0.16}_{-0.03 }$
 \\
A3571         &
$303^{+  14}_{ -14}$&
$-24.5$&
$  6.4^{+  1.9}_{ -1.5}$&
$ 19.1^{+ 73.6}_{-15.2}$&
$ 0.00^{+ 0.20}_{- 0.00 }$
 \\
A1795         &
$297^{+  12}_{ -12}$&
$-23.1 $\tablenotemark{a}&
$  6.0^{+  1.7}_{ -1.4}$&
$  3.9^{+ 14.1}_{ -3.0}$&
$  3.7^{+  0.3}_{ -0.1 }$
 \\
A2029         &
$359^{+  12}_{ -12}$&
$-24.1 $\tablenotemark{a}&
$ 12.3^{+  3.7}_{ -3.0}$&
$ 11.6^{+ 44.0}_{ -9.2}$&
$  7.2^{+  0.6}_{ -1.4 }$
 \\
A2052         &
$195^{+   8}_{  -8}$&
$-22.9$&
$  1.2^{+  0.3}_{ -0.3}$&
$  3.1^{+ 11.3}_{ -2.4}$&
$ 0.68^{+ 0.21}_{-0.01 }$
 \\
MKW3s         &
\nodata &
$-23.1$&
\nodata &
$  4.0^{+ 14.6}_{ -3.2}$&
$ 0.86^{+ 0.05}_{-0.27 }$
 \\
A2065         &
\nodata &
$-22.5 $\tablenotemark{a}&
\nodata &
$  1.9^{+  6.8}_{ -1.5}$&
 0
 \\
A2063         &
$240^{+  52}_{ -52}$&
$-22.6 $\tablenotemark{a}&
$  2.7^{+  3.0}_{ -1.6}$&
$  2.3^{+  8.1}_{ -1.8}$&
$ 0.16^{+ 0.04}_{-0.12 }$
 \\
A2142         &
\nodata &
\nodata &
\nodata &
\nodata &
$  3.3^{+  1.1}_{ -2.9 }$
 \\
A2199         &
$309^{+   7}_{  -7}$&
$-24.0$&
$  6.9^{+  1.7}_{ -1.5}$&
$ 10.7^{+ 40.5}_{ -8.5}$&
$ 1.00^{+ 0.14}_{-0.04 }$
 \\
A2204         &
\nodata &
\nodata &
\nodata &
\nodata &
$ 15.7^{+  3.8}_{ -2.1 }$
 \\
TriAust       &
\nodata &
\nodata &
\nodata &
\nodata &
$ 0.00^{+ 0.03}_{- 0.00 }$
 \\
A2244         &
\nodata &
\nodata &
\nodata &
\nodata &
$  2.1^{+  0.5}_{ -2.1 }$
 \\
A2256         &
$370^{+   9}_{  -9}$&
\nodata &
$ 13.6^{+  4.0}_{ -3.3}$&
\nodata &
 0
 \\
Ophi          &
\nodata &
\nodata &
\nodata &
\nodata &
$ 0.83^{+ 0.28}_{-0.83 }$
 \\
A2255         &
$285^{+  30}_{ -30}$&
\nodata &
$  5.2^{+  2.7}_{ -1.9}$&
\nodata &
 0
 \\
A2319         &
\nodata &
$-23.8 $\tablenotemark{a}&
\nodata &
$  8.5^{+ 31.8}_{ -6.7}$&
 0
 \\
Cyg-A         &
\nodata &
$-23.0 $\tablenotemark{a}&
\nodata &
$  3.3^{+ 12.0}_{ -2.6}$&
$  4.2^{+  0.4}_{ -0.6 }$
 \\
A3667         &
\nodata &
$-23.6 $\tablenotemark{a}&
\nodata &
$  6.8^{+ 25.3}_{ -5.4}$&
 0
 \\
A2597         &
$206^{+  56}_{ -56}$&
$-22.6 $\tablenotemark{a}&
$  1.5^{+  2.3}_{ -1.1}$&
$  2.2^{+  7.8}_{ -1.7}$&
$  3.9^{+  0.7}_{ -1.2 }$
 \\
Klem44        &
$205^{+  11}_{ -11}$&
$-22.9$&
$  1.5^{+  0.4}_{ -0.3}$&
$  2.9^{+ 10.5}_{ -2.3}$&
$ 0.43^{+ 0.15}_{-0.10 }$
 \\
A4059         &
$304^{+  49}_{ -49}$&
$-23.6$&
$  6.5^{+  5.3}_{ -3.3}$&
$  6.5^{+ 24.2}_{ -5.2}$&
$ 0.61^{+ 0.11}_{-0.11 }$
 \\
\enddata
\tablenotetext{a}{The $R$-band magnitude is determined from the $B$-band
 magnitude by $B-R=1.67$.}

\end{deluxetable}

\clearpage

\begin{deluxetable}{ccccccccc}
\tabletypesize{\scriptsize}
\tablecaption{Fitting Parameters. \label{tab:beta}}
\tablewidth{0pt}
\tablehead{
\colhead{Cluster} & \colhead{$z$}   & 
\colhead{$\rho_{\rm gas,0,1}$}   & \colhead{$\rho_{\rm gas,0,2}$} &
\colhead{$r_{c1}$} & \colhead{$r_{c2}$} &
\colhead{$\beta_1$}  & \colhead{$\beta_2$} 
& \colhead{ref\tablenotemark{a}} \\
\colhead{} & \colhead{}   & 
\colhead{($10^{-27}\rm g\; cm^{-3}$)} &
\colhead{($10^{-27}\rm g\; cm^{-3}$)} &
\colhead{(kpc)} & \colhead{(kpc)}  &
\colhead{} & \colhead{}  & \colhead{} 
}
\startdata
A85           &
 0.0556&
67.6&
 6.53&
  36&
 259&
 0.584&
 0.711&
RB2
 \\
A119          &
 0.0440&
\nodata &
 3.18&
\nodata &
 353&
\nodata &
 0.662&
MME
 \\
A262          &
 0.0161&
85.7&
 6.49&
  10&
  97&
 0.556&
 0.556&
MME
 \\
AWM7          &
 0.0172&
40.7&
 9.05&
  23&
 141&
 0.678&
 0.678&
MME
 \\
A399          &
 0.0715&
\nodata &
 4.88&
\nodata &
 333&
\nodata &
 0.713&
RB1
 \\
A401          &
 0.0748&
12.8&
 1.16&
 195&
 887&
 0.696&
 0.952&
RB2
 \\
A3112         &
 0.0750&
128&
22.4&
  20&
  90&
 0.608&
 0.614&
RB2
 \\
A426          &
 0.0183&
126&
 7.25&
  41&
 290&
 0.748&
 0.748&
MME
 \\
2A~0335$+$096 &
 0.0349&
126&
16.8&
  35&
 110&
 0.978&
 0.680&
RB2
 \\
A3158         &
 0.0590&
\nodata &
 9.42&
\nodata &
 193&
\nodata &
 0.657&
MME
 \\
A478          &
 0.0900&
109&
19.0&
  36&
 158&
 0.652&
 0.687&
RB2
 \\
A3266         &
 0.0594&
\nodata &
 5.40&
\nodata &
 364&
\nodata &
 0.744&
MME
 \\
A496          &
 0.0328&
108&
 5.60&
  17&
 200&
 0.551&
 0.733&
RB2
 \\
A3391         &
 0.0531&
12.6&
 4.41&
  34&
 225&
 0.574&
 0.634&
RB2
 \\
A576          &
 0.0381&
\nodata &
 3.63&
\nodata &
 287&
\nodata &
 0.825&
RB1
 \\
PKS~0745$-$191&
 0.1028&
186&
23.9&
  33&
 132&
 0.642&
 0.653&
RB2
 \\
A644          &
 0.0704&
17.9&
 2.93&
 145&
 323&
 0.733&
 0.765&
RB2
 \\
A754          &
 0.0528&
\nodata &
 6.63&
\nodata &
 269&
\nodata &
 0.614&
MME
 \\
HYD-A         &
 0.0538&
107&
15.2&
  35&
 146&
 0.924&
 0.738&
RB2
 \\
A1060         &
 0.0114&
18.3&
 6.87&
  31&
 118&
 0.703&
 0.703&
MME
 \\
A1367         &
 0.0216&
\nodata &
 2.87&
\nodata &
 260&
\nodata &
 0.607&
MME
 \\
Virgo         &
 0.0037&
252&
21.4&
   2&
  21&
 0.420&
 0.470&
M  
 \\
Cent          &
 0.0103&
157&
 7.10&
   9&
  99&
 0.569&
 0.569&
MME
 \\
Coma          &
 0.0232&
\nodata &
 7.14&
\nodata &
 279&
\nodata &
 0.705&
MME
 \\
A1644         &
 0.0474&
\nodata &
 5.03&
\nodata &
 219&
\nodata &
 0.579&
RB1
 \\
A3532         &
 0.0539&
 6.39&
 1.70&
 154&
 638&
 0.811&
 1.234&
RB2
 \\
A1650         &
 0.0845&
\nodata &
 9.73&
\nodata &
 209&
\nodata &
 0.704&
RB1
 \\
A1651         &
 0.0860&
\nodata &
19.8&
\nodata &
 119&
\nodata &
 0.616&
MME
 \\
A1689         &
 0.1840&
62.4&
11.0&
  79&
 311&
 0.808&
 0.874&
RB2
 \\
A1736         &
 0.0460&
\nodata &
 2.92&
\nodata &
 273&
\nodata &
 0.542&
RB1
 \\
A3558         &
 0.0480&
\nodata &
11.3&
\nodata &
 142&
\nodata &
 0.548&
MME
 \\
A3562         &
 0.0499&
\nodata &
13.4&
\nodata &
  71&
\nodata &
 0.470&
MME
 \\
A3571         &
 0.0397&
\nodata &
18.0&
\nodata &
 126&
\nodata &
 0.610&
MME
 \\
A1795         &
 0.0616&
73.2&
 6.30&
  50&
 299&
 0.690&
 0.873&
RB2
 \\
A2029         &
 0.0767&
117&
20.2&
  31&
 137&
 0.607&
 0.647&
RB2
 \\
A2052         &
 0.0348&
71.8&
16.1&
  44&
 110&
 1.465&
 0.678&
RB2
 \\
MKW3s         &
 0.0450&
44.5&
15.2&
  55&
 118&
 1.215&
 0.692&
RB2
 \\
A2065         &
 0.0722&
\nodata &
 4.53&
\nodata &
 511&
\nodata &
 1.162&
RB1
 \\
A2063         &
 0.0354&
21.0&
 5.79&
  51&
 201&
 0.753&
 0.748&
RB2
 \\
A2142         &
 0.0899&
38.1&
 3.38&
  98&
 628&
 0.668&
 0.975&
RB2
 \\
A2199         &
 0.0302&
65.2&
13.9&
  30&
 118&
 0.663&
 0.663&
MME
 \\
A2204         &
 0.1523&
175&
52.6&
 121&
 126&
 4.503&
 0.651&
RB2
 \\
TriAust       &
 0.0510&
14.1&
 5.06&
 155&
 455&
 0.816&
 0.816&
MME
 \\
A2244         &
 0.0970&
32.6&
 9.13&
  69&
 169&
 0.639&
 0.636&
RB2
 \\
A2256         &
 0.0601&
\nodata &
 6.87&
\nodata &
 358&
\nodata &
 0.828&
MME
 \\
Ophi          &
 0.0280&
33.1&
14.5&
  57&
 193&
 0.705&
 0.705&
MME
 \\
A2255         &
 0.0800&
\nodata &
 3.92&
\nodata &
 434&
\nodata &
 0.792&
MME
 \\
A2319         &
 0.0564&
 9.56&
 5.53&
 473&
 558&
 2.602&
 0.812&
RB2
 \\
Cyg-A         &
 0.0561&
\nodata &
294&
\nodata &
  11&
\nodata &
 0.472&
MME
 \\
A3667         &
 0.0560&
\nodata &
 8.27&
\nodata &
 190&
\nodata &
 0.541&
MME
 \\
A2597         &
 0.0852&
120&
 6.70&
  33&
 199&
 0.689&
 0.782&
RB2
 \\
Klem44        &
 0.0283&
34.2&
 3.59&
  38&
 187&
 0.589&
 0.712&
RB2
 \\
A4059         &
 0.0460&
27.6&
 3.06&
  58&
 297&
 0.659&
 0.857&
RB2
 \\
\enddata

\tablenotetext{a}{References--RB1: one $\beta$-model fits by
\citet{rei02}, RB2: clusters in the catalogue of \citet{rei02} fitted
with two $\beta$ models, MME: \citet{moh99}, and M: \citet{mat02}}
\end{deluxetable}

\clearpage

\clearpage

\end{document}